\documentclass[a4paper,12pt]{article}
\usepackage{graphicx}
\usepackage{amssymb}
\usepackage{epstopdf}
\usepackage{bm}
\usepackage[a4paper, total={6.5in, 9in}]{geometry}
\usepackage{hyperref}
\usepackage{xcolor}
\newtheorem{proposition}{Proposition}
\newtheorem{remark}{Remark}

\title{\LARGE Optimal VWAP execution under transient price impact}
\author{Alexander Barzykin and Fabrizio Lillo\footnote{Barzykin: HSBC  FX eRisk, Global Markets, HSBC Bank Plc., UK. 
\href{falexander.barzykin@hsbc.com}{alexander.barzykin@hsbc.com}. Lillo: Dipartimento di Matematica, Universit\`a di Bologna, Italy. \href{fabrizio.lillo@unibo.it}{fabrizio.lillo@unibo.it}.  We thank Jeanine Baumert (who participated to the early stage of this project) and James Newbury for useful discussions. }}
\begin{document}
%\newgeometry{total={7in, 10.5in}}
\maketitle

\abstract{We solve the problem of optimal liquidation with Volume-Weighted Average Price (VWAP) benchmark when the market impact is linear and transient. Our setting is indeed more general as it considers the case when the trading interval is not necessarily coincident with the benchmark interval: Implementation Shortfall and Target Close execution are shown to be particular cases of our setting.  We find explicit solutions in continuous and discrete time considering risk averse investors having a CARA utility function.  Finally, we show that, contrary to what is observed for Implementation Shortfall, the optimal VWAP solution contains both buy and sell trades also when the decay kernel is convex.  
\vspace{0.1cm}}

\section{Introduction}

%\subsection{Literature review}

Optimal execution is becoming a hot field in market microstructure and mathematical finance. The reason is that, with the electronification and fragmentation of financial markets, the execution of a trade requires  advanced infrastructure and sophisticated knowledge on how the trading orders affect prices. In a situation where the available liquidity at the best price is vanishingly small, the best strategy to minimize costs is to split the order in may trades to be executed sequentially, as postulated well before the modern markets by Kyle \cite{kyle}. For this reason, starting from the seminal papers of Bertsimas and Lo \cite{Bertsimas} and Almgren and Chriss \cite{Almgren2}, many contributions have been proposed to find solve the optimal execution problem (for extensive reviews, see \cite{GatheralSchied,GueantBook,Cartea}).

The problem setting crucially depends on two elements: (i) the market impact model and (ii) the benchmark criterion. Concerning the first one, empirical literature \cite{Bouchaud,Moro,Zarinelli,Bouchaud2} has documented that the assumption of permanent and fixed market impact postulated in the original papers \cite{Bertsimas,Almgren2} is not observed. On the contrary, market impact is {\it transient}, i.e. impact is strongest immediately after being triggered and then it decays in time, typically very slowly. This evidence has lead to the development of new models, the most famous one being the Transient Impact Model (TIM, earlier known as the propagator model) \cite{Bouchaud}, whose continuous time version was proposed in \cite{Gatheral}. Other approaches, which model in more details the limit order book \cite{Obizhaeva,Fruth,Donier} behave in some limit as the TIM model\footnote{But of course they deviate in other aspects, thus what is optimal for the TIM is not necessarily so for these other order book models.}.  Originally developed for modeling market impact, the TIM has immediately become subject of intense studies for the optimal execution problem. Refs. \cite{Bouchaud2,Busseti,Gatheral2,curatonew} constitute an incomplete list.

All the above papers, however, solve the optimal execution problem under the assumption that the benchmark criterion is the Implementation Shortfall (IS) or Arrival Price. For a sell order, this means that the execution tries to maximize the risk adjusted difference between the expected revenue from the proceeds and the value of the order marked to market just before the execution starts. Despite being very used in academic literature, the IS is probably not the most used benchmark in the industry. More common alternatives are the Volume Weighted Average Price (VWAP)\footnote{The frequently used Time Weighted Average Price can be seen as a special case of the VWAP when the market volume is considered constant.}, the Target Close (TC), and the Percentage of Volume (POV). Surprisingly, relatively few studies have considered these benchmarks. Among the exceptions, Refs. \cite{Kato,Konishi,Gueant,Frei} considered the VWAP, but the price impact model used is the one of Almgren and Chris (with or without stochastic market volume). 

To the best of our knowledge, there is no paper studying the optimal execution problem under the TIM when the execution is benchmarked against the VWAP (TWAP) or the TC. In this paper we fill this gap by solving the problem in continuous and in discrete time. We frame a setting where a broker has to sell a quantity of shares in a time window $[0, T]$ and we consider as benchmark the VWAP in an interval $[T_1,T_2] \subseteq [0,T]$. In the next Section we motivate when such situation can arise in practice. Here it is worth noticing that IS, TC, and VWAP are special cases of this general problem. The first one when $T_1=0$ and $T_2 \to T_1$, the second when $T_2=T$ and $T_1\to T_2$ and the last one when $T_1=0$ and $T_2=T$. Interestingly, also the case when  $[T_1,T_2] $ is finite and not coincident with $[0,T]$ (dubbed interval VWAP) is of interest in practice, as detailed below.

The problem setting postulates that the broker has a CARA utility function, which is standard in this kind of problems (see, for example, \cite{GueantBook}). We solve the problem in continuous time, by transforming the problem of maximizing of the utility function to the one of solving an integral equation, similarly to what done in \cite{Gatheral2} for the IS case. The setting of the problem in discrete time is instead useful if more constraints (for example on the maximal participation rate) must be added to the problem. In fact, the maximization of the utility function can be transformed in a quadratic optimization problem, to which additional constraints in a linear or even quadratic form can be added, without changing dramatically the complexity of the problem. Finally, the discrete case allows to add the trading volume of the execution to the benchmark, an addition that can be relevant for very large trades.

The paper is organized as follows. Section \ref{sec:tim} introduces the TIM and its known properties also in relation to the existence of dynamic arbitrage. Section \ref{sec:setting} sets the optimization problem in continuous time and find the equivalence with the integral equation. Moreover, explicit solution in simple cases are given. In Section \ref{sec:distime} we restate the problem in discrete time (discussing the connections with continuous time case), solve it, and present some specific numerical examples. 
Finally. in Section \ref{sec:concl} we draw conclusions and provide suggestions for further work.

\section{The Transient Impact Model}\label{sec:tim}

Any optimal execution problem depends critically on the market impact model, i.e. on how the price reacts to trades of the execution. In this paper we will consider the Transient Impact Model (TIM)
(aka the propagator model) introduced in \cite{Bouchaud} (see also \cite{Bouchaud2}). Originally introduced in transaction (i.e. discrete) time, it has been generalized to continuous time in \cite{Gatheral}. Considering a time interval $[0,T]$ and indicating with $S_t$ the price at time $t$, the evolution of the price under the TIM is 
\begin{equation}\label{eq:tim}
S_t=S_0+\int_0^tf(\dot x_s)G(t-s) ds+\int_0^t\sigma_s dW_s
\end{equation}
where  $\dot x_t~dt >0$ is the amount of shares sold by the considered execution in $[t,t+dt]$, $W_s$ is a Wiener process in a suitable probability space, and volatility $\sigma_s$ is a deterministic function. The function $f$ describes the instantaneous impact of the executed trades on price and in the linear case\footnote{Despite empirical literature suggests a nonlinear behavior of $f$ \cite{Bouchaud2}, models, TIM with nonlinear $f$ appears to admit price manipulation \cite{curatonew}.} considered in this paper, it is
\begin{equation}
f(\dot x_t)=-k\dot x_t
\end{equation}
The function $G(t)$, termed the kernel or propagator of the model, describes the delayed effect of trading on price and  $G(t-s)$ characterizes how a trade 
at time $s$ affects the price at time $t$. Since $G$ is generally observed to be a decreasing function \cite{Bouchaud,taranto}, impact is of  {\it transient} nature in this model. By contrast, other models of market impact, such as the one of Almgren and Chriss \cite{Almgren2}, postulate a permanent impact (i.e. a constant $G$) plus a temporary impact affecting only costs.

A significant part of the literature has considered the problem of price manipulation and dynamic arbitrage under different market impact models \cite{Gatheral,Alfonsi3,GatheralSchied,Gatheral2,curatonew,Schneider}. An impact model admits 
price manipulation when there exists a round trip strategy leaving some profit on expectation \cite{GatheralSchied}. The model, instead, admits transaction triggered price manipulation if the expected revenues of a sell (buy) program can be increased by intermediate buy (sell) trades" \cite{Alfonsi3}.  It is possible to show that the absence of transaction-triggered price manipulation implies the absence of price manipulation \cite{GatheralSchied}.

When considering the TIM as a market impact model, a series of conditions for the absence of market manipulation have been derived (see \cite{GatheralSchied}). In particular, when the function $f$ in (\ref{eq:tim}) is linear, the convexity of $G$ is sufficient to guarantee the absence of transaction triggered price manipulation. This result is however obtained when considering Implementation Shortfall as the function to minimize. In this paper we show that when the objective is different (in our case the VWAP), convexity of $G$ is not anymore sufficient and indeed optimal solutions for a sell program contain both buy and sell trades.

\section{Problem setting in continuous time}\label{sec:setting}
%A broker receives from a client (typically an asset manager) an order to sell a quantity $x_0>0$ shares and does it in a time window $[0,T]$, termed the {\it trading interval}. However, in full generality, the benchmark price is the VWAP in a time window $[T_1,T_2]\subset [0,T]$, termed the {\it benchmark interval}. We can imagine that the times $T_1$ and $T_2$ are set by the client, whereas the broker is allowed to trade in the larger time window $[0,T]$. {\bf @SASHA: Here I need your input to explain under which conditions the trading interval is different from the benchmark interval, when this difference is allowed and when it is not (because it is a sort of front running), and a possible name for a VWAP execution where the trading and the benchmark intervals are different}.

We consider a general problem setup where a broker has to sell a quantity of $x_0>0$ shares in a time window $[0, T]$, termed the {\em trading interval}, and she is benchmarked against the market VWAP in a time window $[T_1,T_2] \subseteq [0,T]$, termed the {\em benchmark interval}. This formulation covers the majority of standard settings, such as Implementation Shortfall with $T_1=T_2=0$, Target Close with $T_1=T_2=T$, and interval VWAP with $T_1=0$, $T_2=T$. The general situation of $[T_1,T_2] \subset [0,T]$ may arise where the broker has to guarantee the VWAP price in a given time interval but $x_0$ is too large to be traded within this interval due to constraints (e.g., maximum POV). Point in time benchmarks, such as Market Close, provide a limiting scenario, which is typically solved with Target Close algorithm executing before Market Close. Trading after close is possible for certain instruments and may be preferable due to volatility risk reduction. Industry point in time benchmarks are being replaced with interval benchmarks, thus supporting the general formulation.

The normalization condition requires
\begin{equation}
\int_0^T \dot x_t dt = x_0
\end{equation}
even if it is possible that $\int_0^T |\dot x_t| dt > x_0$, i.e. the broker can decide also to buy in the market a part of shares (if this is allowed). 
Let $V_tdt$ be the deterministic market volume traded in $[t,t+dt]$.
The VWAP benchmark is given by
\begin{equation}
VWAP_{T_1}^{T_2}=\frac{\int_{T_1}^{T_2}S_tV_tdt}{\int_0^T V_t dt}=\int_0^T \eta_tS_t dt
\end{equation}
where
\begin{equation}
\eta_t=\frac{V_t}{\int_{T_1}^{T_2} V_s ds} I_{t\in [T_1,T_2]}
\end{equation}
where $I_B$ is the indicator function of the set $B$.

The objective function of the broker is the difference between the cash she is able to obtain from the proceeds in the trading interval and the cash she will give back to the client, equal to $x_0 VWAP_{T_1}^{T_2}$. This difference is of course a random variable, thus we must assume some utility function to model the risk aversion of the broker. To this end let us define the cash process
\begin{equation}
dX_t=\dot x_tS_tdt~~~~~~~~~~~~X_0=0.
\end{equation}
Assuming a CARA risk averse agent, the objective function for a strategy ${\mathbf x}\equiv \{x_t\}_0^T$ is
\begin{equation}\label{eq:opt}
U[{\mathbf x}]={\mathbb E}_0[-\exp(-2\gamma(X_T-x_0VWAP_{T_1}^{T_2}))]
\end{equation}
where $2\gamma$ is the risk aversion parameter. We can now plug in the TIM of Eq. (\ref{eq:tim}) for the dynamics of price and prove the following proposition:

\begin{proposition} Under linear impact, $f(z)=-kz$ with $k>0$, the maximization of the utility function (\ref{eq:opt}) is equivalent to the minimization of the functional
\begin{eqnarray}\label{eq:cost}
C[{\mathbf x}]\equiv 
\frac{1}{2}\int_0^T \int_0^T \dot x_t \dot x_s G(|t-s|)ds~dt-x_0\int_0^T \eta_t dt \int_0^t G(t-s) \dot x_sds \\
+\frac{\gamma}{k}\int_0^T \int_0^T dt~dt'(\dot x_t-x_0\eta_t)(\dot x_{t'}-x_0\eta_{t'})\int_0^{t \wedge t'} \sigma_s^2 ds\nonumber
\end{eqnarray}
\end{proposition}
The proof of this and the following propositions are in the appendix. 
 
In order to find the optimal execution we make use of calculus of variations following the approach of \cite{Gatheral2,LeHalle}. We consider a strategy 
\begin{equation}
dy_s=\delta_{t_2}(ds)-\delta_{t_1}(ds)~~~~~~~~~~0\le t_1 \le t_2 \le T
\end{equation}
corresponding to a instantaneous purchase of one unit at time $t_1$ which is sold instantaneously at time $t_2$.  Indicating with  ${\mathbf x^*}$ the optimal strategy and setting ${\mathbf z}={\mathbf x^*}+\alpha {\mathbf y}$, the integral equation satisfied by the optimal strategy is obtained by setting 
\begin{equation}
\frac{\partial E[C[{\mathbf z}]]}{\partial \alpha} \bigg |_{\alpha=0} =0
\end{equation}
Although it is possible to obtain the integral equation in the general case, in the following we will restrict our attention to the case of a risk neutral investor ($\gamma=0$). In Section \ref{sec:distime} we will explore also the case of a risk averse investor ($\gamma>0$) in discrete time setting. This procedure leads to the following proposition:
\begin{proposition} The strategy $\{x^*_t\}_0^T$ minimizing the functional (\ref{eq:cost}) with $\gamma=0$ satisfies the integral equation
\begin{equation}\label{eq:inteq}
\int_0^T G(|t-s|)dx_s^*-x_0\int_t^T\eta_s G(s-t)ds=\lambda
\end{equation}
 where $\lambda$ is a constant set by the normalization of the total volume traded
 \begin{equation}
 \int_0^T dx^*_s=x_0
 \end{equation}
\end{proposition}

\begin{remark}  \label{rem1}
We remind \cite{Gatheral2} that the optimal execution under TIM when the objective function is the Implementation Shortfall satisfies the equation
\begin{equation}\label{eq:gss}
\int_0^T G(|t-s|)dx_s^*=\lambda
\end{equation}
thus under VWAP objective function there is an additional term $-x_0\int_t^T\eta_s G(s-t)ds$ in the left hand side of the equation. 
\end{remark}

\begin{remark} 
When $T_1=T_2=0$, it is $\eta_t=2\delta(t)$, and the second integral in Eq. \ref{eq:inteq} becomes\footnote{Note that here and in the following we use the convention that $\int_a^b \delta(x-a)f(x)dx= f(a)/2$. Thus, since $\int_0^T \eta_tdt=1$, when $\eta_s$ is a Dirac delta centered either at $t=0$ or at $t=T$, we must include a factor $2$.}
$$
-x_0\int_t^T\delta(s) G(s-t)ds=0
$$
Since $t>0$, the integral equation reduces to Eq. \ref{eq:gss} i.e. the one obtained by Schied et al. \cite{Gatheral2} for the the optimization of the Implementation Shortfall.
\end{remark}

We can use Remark \ref{rem1} to write the solution of the integral equation as the sum of two terms. To this end we introduce the variable
$$
w_s=\dot x_s^*-x_0 \eta_s
$$ 
with $\int_0^T w_s ds=0$ and, by replacing in (\ref{eq:inteq}), we obtain
 \begin{equation}\label{eq:inteq2}
\int_0^T G(|t-s|)w_sds=\lambda-x_0\int_0^t\eta_s G(t-s)ds
\end{equation}
One can write the solution $w_s=w^{(1)}_s+ w^{(2)}_s$ where the second term solves
\begin{equation}\label{eq:inteq3}
\int_0^T G(|t-s|)w^{(2)}_sds=-x_0\int_0^t\eta_sG(t-s)ds
\end{equation}
Setting $x'_0=\int_0^T w^{(2)}_s ds$, the first term solves
$$
\int_0^T G(|t-s|)w^{(1)}_sds=\lambda,~~~~~~~~~~~~\int_0^T w^{(1)}_s ds =-x'_0
$$
which is the equation when the objective function is the IS and the number of shares is $-x'_0$.

\begin{remark}
When $T_1=T_2=T$ (Target Close) it is  $\eta_t=2\delta(t-T)$, the second integral in Eq. \ref{eq:inteq} reduces to
$$
-x_0\int_t^T2\delta(s-T) G(s-t)ds=-x_0G(T-t)
$$
and the integral equation becomes
$$
\int_0^T G(|t-s|)dx_s^*=\lambda+x_0G(T-t)
$$
The solution of this integral equation is $\dot x_s^*=w_s^{(1)}+x_0\delta(T-t)$ with the normalization $\int_0^T w_s^{(1)}ds =x_0/2$. In other words, the optimal schedule under the TC benchmark is the sum of $x_0/2$ shares traded as in the IS case and the remaining $x_0/2$ shares traded at $t=T$.
\end{remark}

\subsection{Explicit Solution for a VWAP when the benchmark interval and the trading interval coincide}
We consider here the case when the benchmark VWAP interval $[T_1,T_2]$ coincides with the trading interval $[0,T]$ and $\eta_t=1/T$, $\forall t \in [0,T]$, i.e. the market volume is constant in the interval (TWAP). 
In this case we are able to find the explicit solution for two different kernels and compare the results with the optimal schedule obtained under different impact models. More general cases will be explored numerically using time discretization in Section \ref{sec:distime}.

\subsubsection{Exponential kernel}
 We consider first the exponential kernel $G(t)=e^{-\rho t}$. It is known that this type of kernel is consistent with the model of Obizhaeva-Wang \cite{Obizhaeva} for the resilience of the order book. We remind that when minimizing  the IS, the solution is to trade a finite fraction $1/(2(1+\rho T))$ instantaneously at times $t=0$ and $t=T$ and a fraction $\rho T/(1+\rho T)$ at constant speed in $(0,T)$. As for any kernel, the IS optimal trading schedule is symmetric with respect to $T/2$. 
 
Defining the trading velocity $v_t\equiv \dot x_t$, it is straightforward to test that the solution of the type
 \begin{equation}
v_t =a_1\delta(t)+b+a_2\delta(t-T)
 \end{equation}
 satisfies (\ref{eq:inteq}), and, by imposing the normalization condition, we obtain the result
 \begin{equation}
 v_t=\frac{x_0}{\rho T(2+\rho T)}\left[2(1+\rho T)\delta(t)+\rho(1+\rho T) -2 \delta (t-T)\right]
 \end{equation}
 Therefore, it is optimal to sell a finite amount at time $t=0$, then selling at a constant rate for the whole interval $[0,T]$ and finally {\it buying} a finite amount at time $t=T$.  

Thus we see that differently from the case of IS, the optimal execution under VWAP (i) is not anymore symmetric around $T/2$ and (ii) allows for transaction triggered price manipulation even when $G$ is convex. As we will see next, these properties also hold for other choices of the kernel function.

\subsubsection{Power law kernel}

It is well known that empirical data show unambiguously that the kernel $G(t)$ behaves as a power law for large lags \cite{Bouchaud,Bouchaud2}: indeed for small tick stocks the power law behavior is observed for all values of $\tau$, while for small tick stocks there is a bump for very small lags \cite{taranto}. For analytical tractability we will consider here the case $G(t)=t^{-\kappa}$ with $\kappa<1$. %Typical empirical values are $\kappa\simeq 0.5$.

To find the optimal solution of the VWAP execution, we use the decomposition leading to (\ref{eq:inteq3}). We remind that in the case of IS the optimal solution for an execution of $-x'_0$ shares is \cite{Bouchaud2,Gatheral2}
\begin{equation}\label{eq:gsssol}
w^{(1)}_t=\frac{-x'_0}{T} \frac{2^\kappa \Gamma\left(1+\frac{\kappa}{2}\right)}{\sqrt{\pi}\Gamma\left(\frac{1+\kappa}{2}\right)}\frac{1}{\left[\frac{t}{T}\left(1-\frac{t}{T}\right)\right]^{(1-\kappa)/2}}
\end{equation}
Note that $w^{(1)}_t$ is always positive in $[0,T]$ and symmetric with respect to $T/2$ (U-shaped), diverging at $t=0^+$ and $t=T^-$.

The function $w^{(2)}_s$ solves the integral equation
 $$
\int_0^T \frac{w^{(2)}_s}{|t-s|^\kappa}ds=- \frac{x_0}{T(1-\kappa)} t^{1-\kappa} \equiv f(t)
 $$
  \medskip
 
This is a generalized Abel integral equation with constant limits \cite{Gakhov,Estrada}. Using Eqs. (2.58) and (2.62) of \cite{Estrada}, the solution can be found as\footnote{Note that the following expression can be used also for finding the optimal execution for a generic $\eta_s\ne 1/T$, since when the kernel is power law, Eq. \ref{eq:inteq} is a generalized Abel integral equation with constant limits.}
\begin{eqnarray}
&&w^{(2)}_t= -\frac{\cos^2\frac{\pi \kappa}{2}}{\pi^2}\frac{1}{[t(T-t)]^{(1-\kappa)/2}} 
{\cal P}\int_0^T\frac{[s(T-s)]^{(1-\kappa)/2}}{s-t}\left(\frac{d}{ds}\int_0^s \frac{f(u)}{(s-u)^{1-\kappa}} du\right)ds   \nonumber\\
&&+\frac{\sin \pi\kappa}{2\pi}\frac{d}{dt}\int_0^t\frac{f(s)}{(t-s)^{1-\kappa}}ds=  \nonumber\\ 
&&=\frac{x_0}{T(1-\kappa)}\times \\
&&\left[ \frac{\cos^2\frac{\pi \kappa}{2}}{\pi^2}\frac{1}{[t(T-t)]^{(1-\kappa)/2}} 
{\cal P}\int_0^T\frac{[s(T-s)]^{(1-\kappa)/2}}{s-t}\left(\frac{d}{ds}\int_0^s \frac{u^{1-\kappa} }{(s-u)^{1-\kappa}} du\right)ds  \right. \nonumber\\
&&\left. -\frac{\sin \pi\kappa}{2\pi}\frac{d}{dt}\int_0^t\frac{s^{1-\kappa}}{(t-s)^{1-\kappa}}ds \right] \nonumber
\end{eqnarray}
where ${\cal P}$ indicates the Cauchy principal value integral. When $\kappa<1$, this expression can be rewritten as
\begin{eqnarray}
&&w^{(2)}_t= \frac{x_0}{T} \pi \csc \pi\kappa \times \\
&& \left[ \frac{\cos^2\frac{\pi \kappa}{2}}{\pi^2} \frac{1}{[t/T(1-t/T)]^{(1-\kappa)/2}} {\cal P} \int_0^1 \frac{[z(1-z)]^{(1-\kappa)/2}}{z-t/T} dz-\frac{\sin \pi\kappa}{2\pi}\right]
\end{eqnarray}
Since
\begin{eqnarray}
&&{\cal P} \int_0^1 \frac{[z(1-z)]^{(1-\kappa)/2}}{z-t/T} dz =\\
&&\frac{2^{\kappa-1}\sqrt{\pi}\Gamma\left(\frac{1-\kappa}{2}\right)~_{2}F_1(1,-1+\kappa,(1+\kappa)/2; t/T)}{\Gamma\left(1-\frac{\kappa}{2}\right)}-\frac{\pi \tan \frac{\kappa \pi}{2}}{{[t/T(1-t/T)]^{(\kappa-1)/2}} }, \nonumber
\end{eqnarray}
where $_{2}F_{1}$ is the hypergeometric function, the solution is
\begin{eqnarray}
w^{(2)}_t=\frac{x_0}{T}\left[-1+\frac{2^{\kappa-2}\sqrt{\pi}\csc(\kappa\pi/2)}{\Gamma\left(1-\frac{\kappa}{2}\right)\Gamma\left(\frac{1+\kappa}{2}\right)}\frac{_2F_1(1,-1+\kappa,(1+\kappa)/2; t/T)}{[t/T(1-t/T)]^{(1-\kappa)/2}}\right]
\end{eqnarray}
By direct integration
$$
x'_0=\int_0^T w^{(2)}_t dt= -\frac{x_0}{2}
$$
This value is used in the expression for $w^{(1)}$ and finally we obtain for the trading velocity $v_t\equiv \dot x_t$
\begin{eqnarray}
v_t=\frac{x_0}{T} \frac{2^{\kappa-2}\sqrt{\pi}\csc(\frac{\kappa\pi}{2})}{\Gamma\left(1-\frac{\kappa}{2}\right)\Gamma\left(\frac{1+\kappa}{2}\right)}\frac{\left[\kappa+~_2F_1\left(1,-1+\kappa,\frac{1+\kappa}{2}; \frac{t}{T}\right)\right]}{\left[\frac{t}{T}\left(1-\frac{t}{T}\right)\right]^{(1-\kappa)/2}}
\end{eqnarray}
Since $_2F_1\left(1,-1+\kappa,\frac{1+\kappa}{2}; 0\right)=1$, $v_t$ diverges positively when $t\to 0^+$ (as in the IS case), whereas the conditions $_2F_1\left(1,-1+\kappa,\frac{1+\kappa}{2}; 1\right)=-1$ and $\kappa<1$ imply that $v_t$ diverges {\it negatively} for $t\to T^-$. This means that in a VWAP sell execution it is optimal to buy toward the end of the trading period and thus that this strategy allows for transaction triggered price manipulation even when $G$ is convex. Finally, it is interesting to observe that in the limit $\kappa \to 1$, the optimal schedule is $v_t=x_0/T$, i.e. to trade at constant speed.

\begin{figure}[t]
	\centering
	\includegraphics[width=0.45\textwidth]{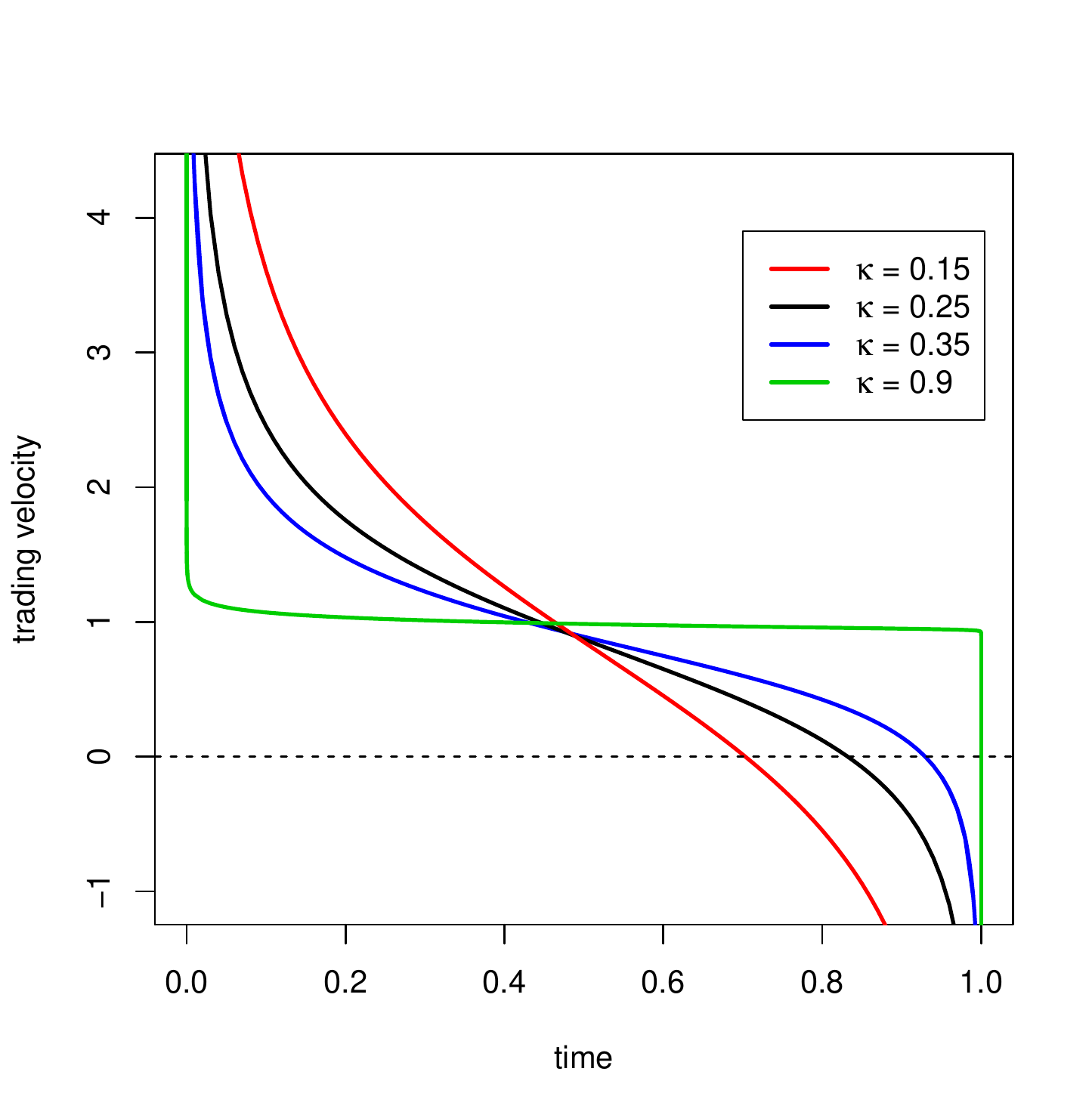}
	\includegraphics[width=0.45\textwidth]{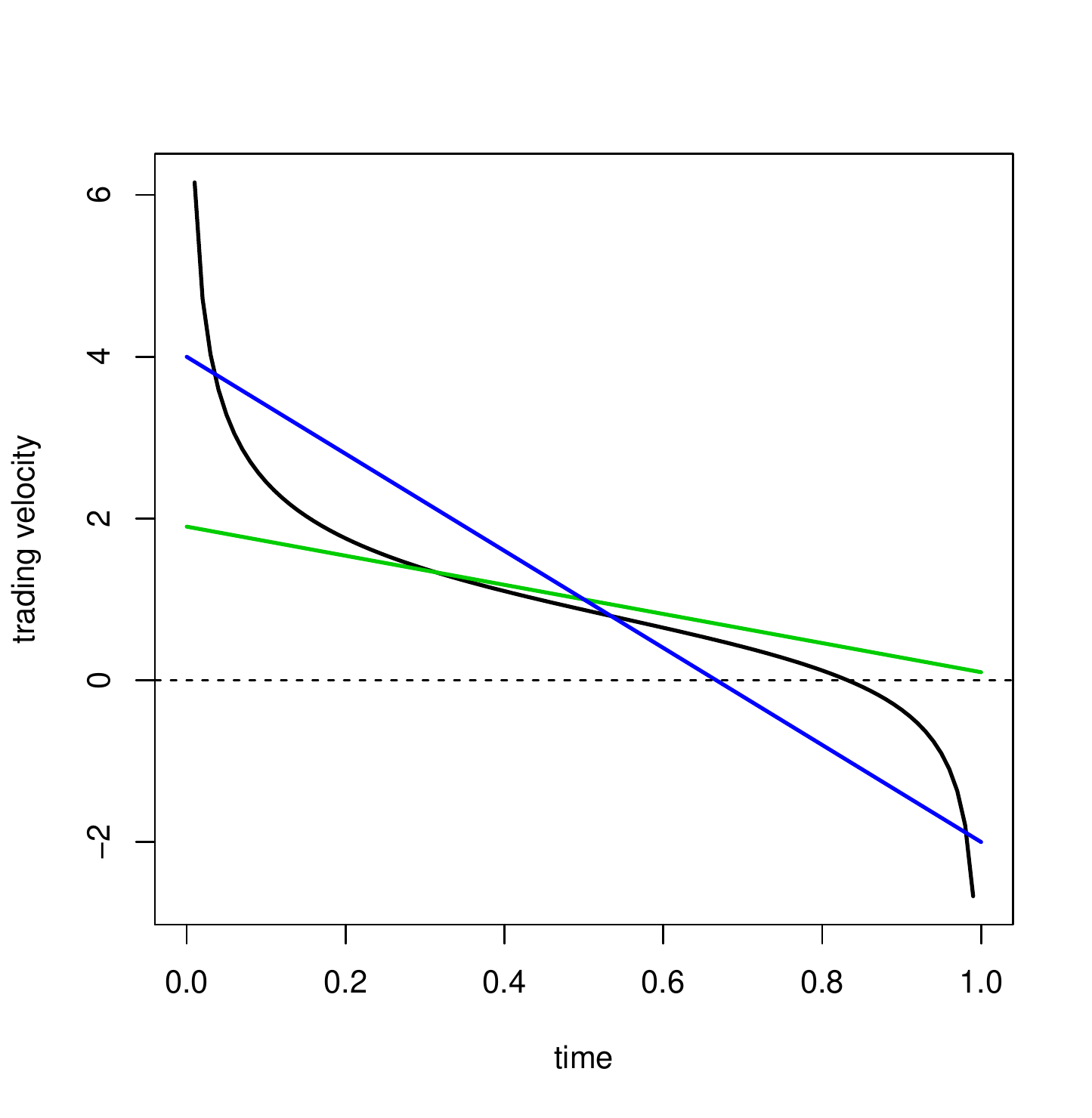}
	\caption{Optimal trading schedule for a VWAP with benchmark interval coincident with the trading interval $[0,1]$. The price is without drift and the broker is risk neutral. Left panel repors four schedules for different values of the exponent $\kappa$ of the kernel. Right panel. The black line is the optimal solution for TIM with $\kappa=0.25$. The green ($\beta=0.9$) and blue ($\beta=3$) lines are the solutions under permanent linear price impact (as in Almgren-Chriss) obtained in \cite{Gueant}.}\label{fig:powlawcont}
\end{figure}

For illustrative purposes we consider the case of $x_0=1$ and $T=1$. The left panel of Figure \ref{fig:powlawcont} shows the optimal schedule for $\kappa=0.15, 0.25, 0.35, 0.9$. Clearly this solution is not symmetric w.r.t. $T/2$. Moreover, qualitatively similarly to the case of exponential kernel, the optimal solution is to sell very intensely around $t=0$, then to sell at a lower speed, and to {\it buy} very intensely when getting close to $t=T=1$. When $\kappa$ is small, the region of negative trading velocity becomes larger. On the contrary, as mentioned above, when $\kappa$ becomes close to one, the optimal schedule becomes close to a constant with a positive and negative peak at $t=0$ and $t=T$, respectively.

\begin{remark} It is interesting to compare our solution with $\kappa=0.25$ with the one obtained in \cite{Gueant} for a linear permanent impact model \'a la Almgren-Chriss. Assuming a linear permanent impact with constant $k$ and a quadratic temporary impact with constant $\eta$, \cite{Gueant} finds that the optimal solution for a risk neutral agent is
$$
v_t=x_0[(\beta+1)-2\beta t]
$$
where $\beta=kV/4\eta$ and $V$ is the market volume. Figure \ref{fig:powlawcont} shows two solutions corresponding to $\beta=0.9$ (green) and $\beta=3$ (blue). Qualitatively, also in these solutions it is optimal to trade faster at the beginning of the interval and, in the case of large $\beta$, (i.e. small temporary impact) it is optimal to buy back a part toward the end. 
\end{remark}

\section{Solution in discrete time}\label{sec:distime}

In this Section we derive the solution of the optimal VWAP by using a discrete time framework. This setting will allow to obtain explicit solutions and to explore also the role of additional constraints (for example the requirement that in a sell program no buying is allowed). The discrete time setting can be applied at three different levels: (i) express the cost function (\ref{eq:cost}) in discrete time and solve the optimization; (ii) use discrete time to obtain a quadrature of the integral equation (\ref{eq:inteq3}); (iii) write the TIM (\ref{eq:tim}) in discrete time, derive the corresponding cost, and then minimize it. It is worth noticing that the three procedures do not give exactly the same result, however if the time intervals used in the discretization are sufficiently small, the differences become negligible. In the following we will consider approach (iii) and we will briefly discuss the difference with approach (i).

Let us divide the interval $[0,T]$ in $N$ equal intervals and define $\tau=T/N$. The strategy is now a vector ${\mathbf x}=(x_1,....,x_N)'$, where $x_i$ is the amount of shares traded in interval $i$, i.e. for $t\in[(i-1)\tau, i\tau]$.  The price dynamics of a sell execution in discrete time is
\begin{equation}
S_\ell=S_0- k \sum_{i=1}^\ell G(\ell-i) x_i+ \tau^{1/2} \sum_{i=1}^\ell \epsilon_i ~~~~~\ell=\{0,...,N\}
\end{equation}
which can be rewritten in vector form as
\begin{equation}
{\mathbf S} = S_0 {\mathbf 1} -k G {\mathbf x}+ \tau^{1/2}L {\bm \epsilon}
\end{equation}
where ${\mathbf S}=(S_1,...,S_N)'$, ${\mathbf 1}=(1,...,1)'$, $L$ is the lower triangular matrix of ones (i.e. $L_{ij}=1$ if $i\ge j$, zero otherwise), and $G$ is the lower triangular matrix such that $G_{ij}=G[\tau(i-j)]$ if $i\ge j$ and zero otherwise. 
Finally ${\bm \epsilon}\sim {\mathcal N}({\bm \mu}, \Sigma)$ is a Gaussian random vector describing the price dynamics without execution. 
Even if in the following we will focus mainly on ${\bm \mu}={\mathbf 0}$ and $\Sigma=diag (\sigma^2_i)$, we will provide solutions in the presence of drift and correlated returns.
The cash amount at the end of the period is $X_N={\mathbf x}' {\mathbf S}$. 

In full generality, we consider a VWAP benchmark between $t=T_1$ and $t=T_2$, corresponding to $\ell_1=\lfloor NT_1/T\rceil$ $\ell_2=\lfloor NT_2/T\rceil $ are the rounding to the nearest integer giving the initial and final trading intervals. We indicate $B=\{\ell\in{\mathbb N}: \ell_1\le \ell \le \ell_2\}$ and we introduce a vector ${\bm \eta}$ with components 
\begin{equation}\label{bench}
\eta_\ell=\frac{V_\ell}{||{\bm \eta}||_1} I_{\ell\in B}
\end{equation}
where $V_\ell$ is the market volume traded in interval $\ell$ and $I$ is the indicator function\footnote{For the moment we are neglecting our trading on the benchmark, see the subsection 4.2.}.
The benchmark is $x_0{\bm\eta}' {\mathbf S}$ and the normalization ensures that ${\mathbf 1}' {\bm \eta}=1$. The utility function is  ${\cal U}[({\mathbf x}-x_0{\bm \eta})'  {\mathbf S}]$ and, using the Gaussian assumption under CARA utility function with risk aversion $2\gamma$, the expected utility is
\begin{equation}\label{eq:ut}
U[{\mathbf x}]={\mathbb E}_0[ ({\mathbf x}-x_0{\bm \eta})' {\mathbf S}] -\gamma {\mathbb V}_0[ ({\mathbf x}-x_0{\bm \eta})' {\mathbf S}]
\end{equation}
In the Appendix we prove the following proposition.

\begin{proposition} \label{prop:discrete}
Under CARA utility function with risk aversion $2\gamma$, the optimal VWAP execution, which maximizes the expected utility (\ref{eq:ut}), is the solution of the quadratic optimization
$$
\min_{\mathbf x}\left[{\mathbf x}' A {\mathbf x} - {\mathbf b}' {\mathbf x}\right]~~~~~~~~   s.t. ~~~ {\mathbf 1}' {\mathbf x}=x_0
$$
where 
\begin{eqnarray}
&A&=kG+\gamma\tau L\Sigma L' \label{eq:param1} \\
&{\mathbf b}'&= kx_0{\bm \eta}' G + 2\gamma\tau x_0 {\bm \eta}'L\Sigma L'+\tau^{1/2}{\bm \mu}' L'\label{eq:param2}
\end{eqnarray} 
Moreover, the matrix $A$ is positive definite if $\Sigma$ is positive definite. Thus the solution of the quadratic optimization exists and is unique.
\end{proposition}

Since the problem can be recast in a quadratic optimization form, several additional constraints can be added without affecting the difficulty of the problem. 
For example, it is possible to add the constraint that all the trades have the same sign, e.g. no buys in a sell execution ($x_i\ge 0, \forall i$), 
or a constraint on the maximal trading speed ($|x_i|\le x_{max}, \forall i$).

\begin{remark}
Note that when $T_1=1$ and $T_2=N$, one does not obtain the same solution derived  by discretizing the cost function. This is because here we have discretized the impact model and not the cost. The difference between the two solutions is due to the diagonal terms of $G$ which when discretizing the cost are half than those obtained by discretizing the impact model. The difference between the strategies is however small and tends to zero when $N\to \infty$.
%, see right panel of Fig. \ref{vwapt}.  
\end{remark}

\subsection{Numerical results}

In this section we explore the optimal solutions under different parameter choices. In all analyses we will set $\tau=1$, $T=N=50$, $k=1$, and $x_0=1000$. Moreover we choose a power law kernel $G_{ij}=\frac{1}{2+|i-j|^\kappa}$ with $\kappa=0.5$. Finally we assume a flat market volume profile, i.e.  $V_\ell=const$.

\begin{figure}[t] 
	\centering
	\includegraphics[width=0.5\textwidth]{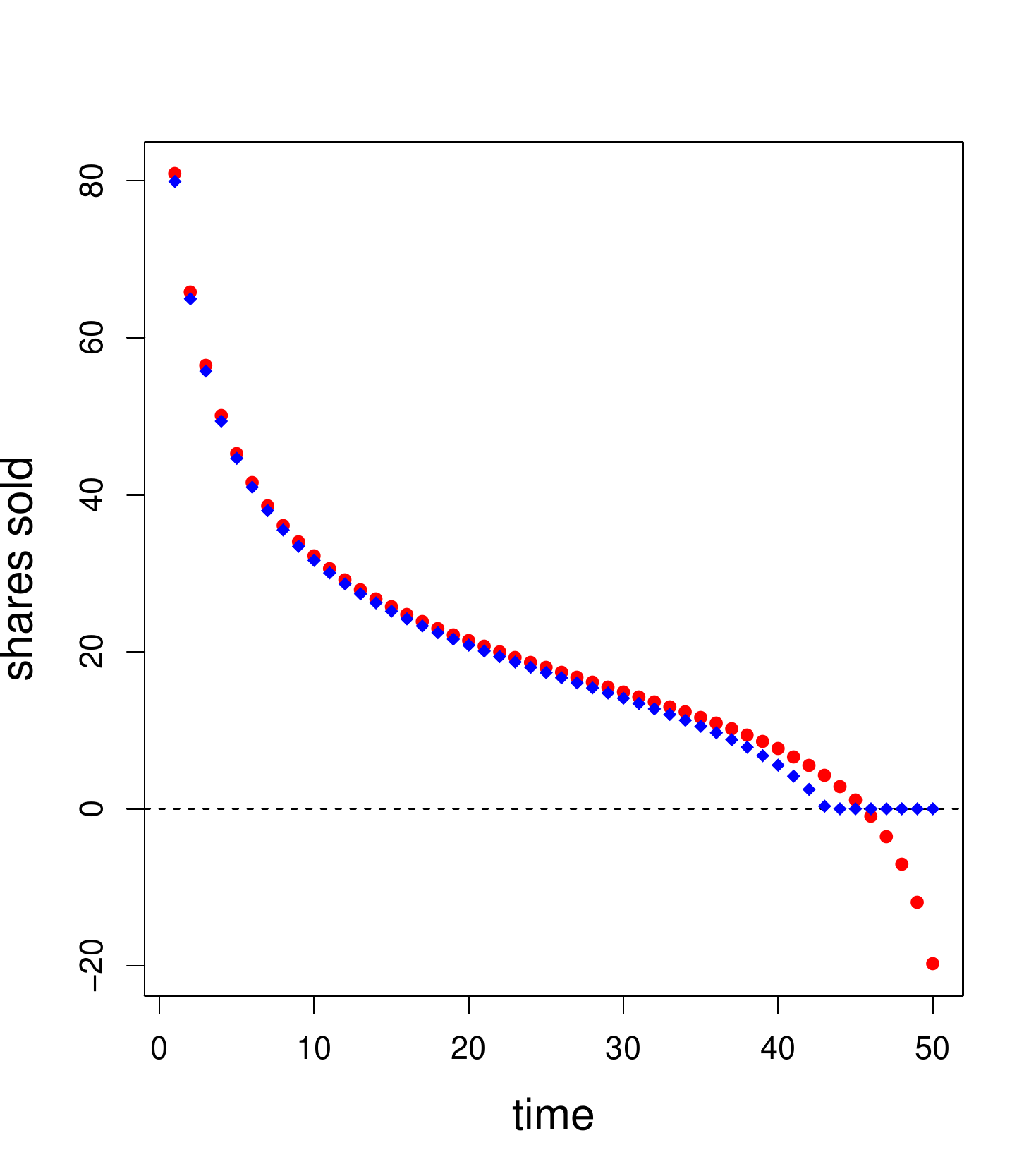}
	\caption{Optimal trading schedule for a VWAP with benchmark interval coincident with the trading interval. The price is without drift and the broker is risk neutral. The red dots refer to the unconstrained case, while the blue ones to the case with a constraint on the non-negativity of trades (no buys for a sell execution).}
	\label{fig:vwapbase}
\end{figure}

Figure \ref{fig:vwapbase} shows the baseline case where $\gamma=0$ (i.e. a risk neutral broker), ${\bm \mu}={\mathbf 0}$ (no drift), $T_1=0$ and $T_2=T$. The red dots refer to the unconstrained problem, while the blue ones to the case with constraint $x_i\ge 0, \forall i$.  The first one is similar to the solution in continuous time shown in Fig. \ref{fig:powlawcont} with negative positions (buys) toward the end of the execution, while in the second one the negative $x_i$ are essentially capped to zero. In both cases the value of the expected utility, which in the risk neutral case corresponds to the expected cash minus the VWAP, is positive indicating a liquidation value larger than the VWAP, thus a net profit for the broker.

\begin{figure}[t] 
	\centering
	\includegraphics[width=0.45\textwidth]{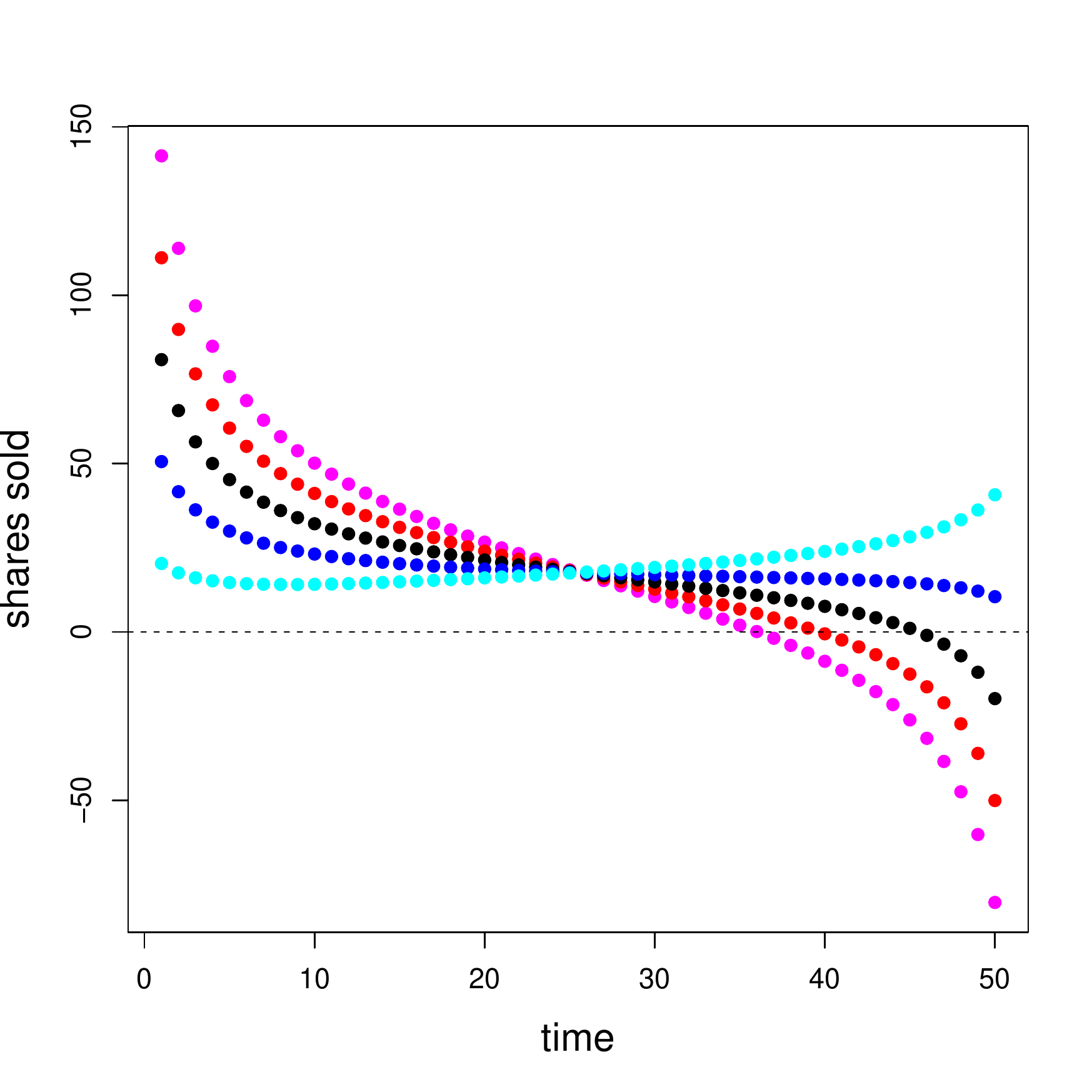}
	\includegraphics[width=0.45\textwidth]{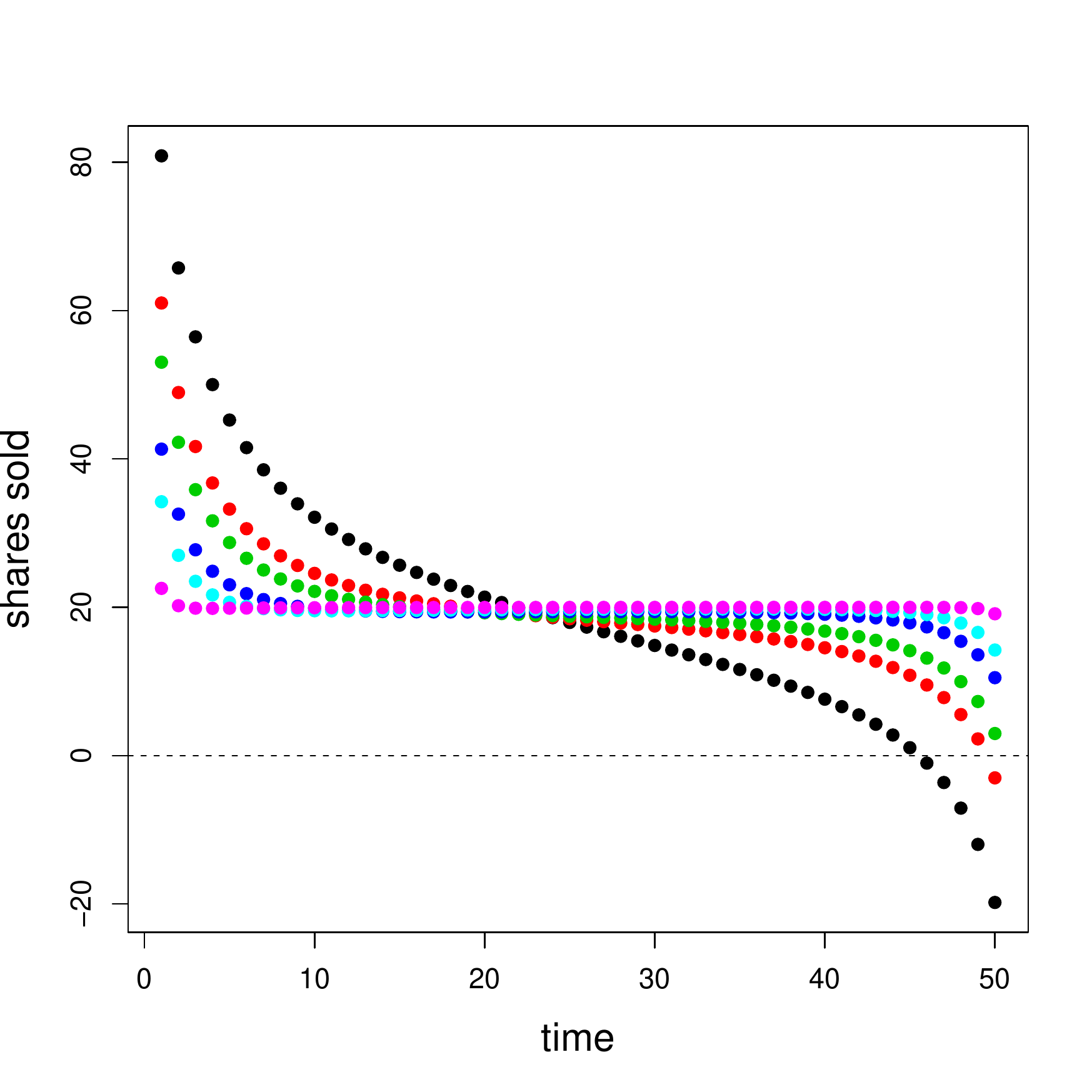}
	\caption{Left. Optimal VWAP schedule for a sell order by a risk neutral broker for different values of the price drift $\mu_i=4$ (cyan), $\mu_i=2$ (blue)  $\mu_i=-2$ (red), and $\mu_i=-4$ (magenta). Black dots refer to the driftless benchmark case. Right. Optimal VWAP schedule for a risk averse broker under driftless price. The values of the risk aversion parameter $\gamma$ are $0$ (black), $0.5$ (red), $1$ (green), $3$ (blue), $7$ (cyan), $100$ (magenta). In both panels the benchmark interval is coincident with the trading interval.}
	\label{fig:vwapdriftrisk}
\end{figure}

We then consider the role of drift and risk aversion. Left panel of Fig. \ref{fig:vwapdriftrisk} shows the optimal solution for risk neutral agents and different values of the (constant) drift. Blue (red) lines refer to positive (negative) drift, while the black line refers to the driftless benchmark case. As is intuitive, when the drift is positive (negative), it is optimal to delay (anticipate) the sale of the shares. The right panel of Fig. \ref{fig:vwapdriftrisk} shows the optimal solution in the driftless case for different risk aversion parameter $\gamma$. We set $\Sigma=diag (\sigma^2_i)$ with a constant volatility $\sigma_i^2=0.01$ $\forall i$. 

\begin{remark} 
It is interesting to note that for very large risk aversion, the optimal trading profile becomes flat, i.e. if the broker cares only about the variance of the profit, the optimal choice is to trade at constant speed, which, under the assumption of constant market volume $V_\ell$, means fixed percentage of volume (like in a POV strategy). Note that this is different to what happens under IS benchmark \cite{Busseti}, since in this case the risk neutral U-shape becomes asymmetric and the strategy is front loaded (i.e. more trading at the beginning of the execution than at the end) 
\end{remark}

\begin{figure} 
	\centering
	\includegraphics[width=0.45\textwidth]{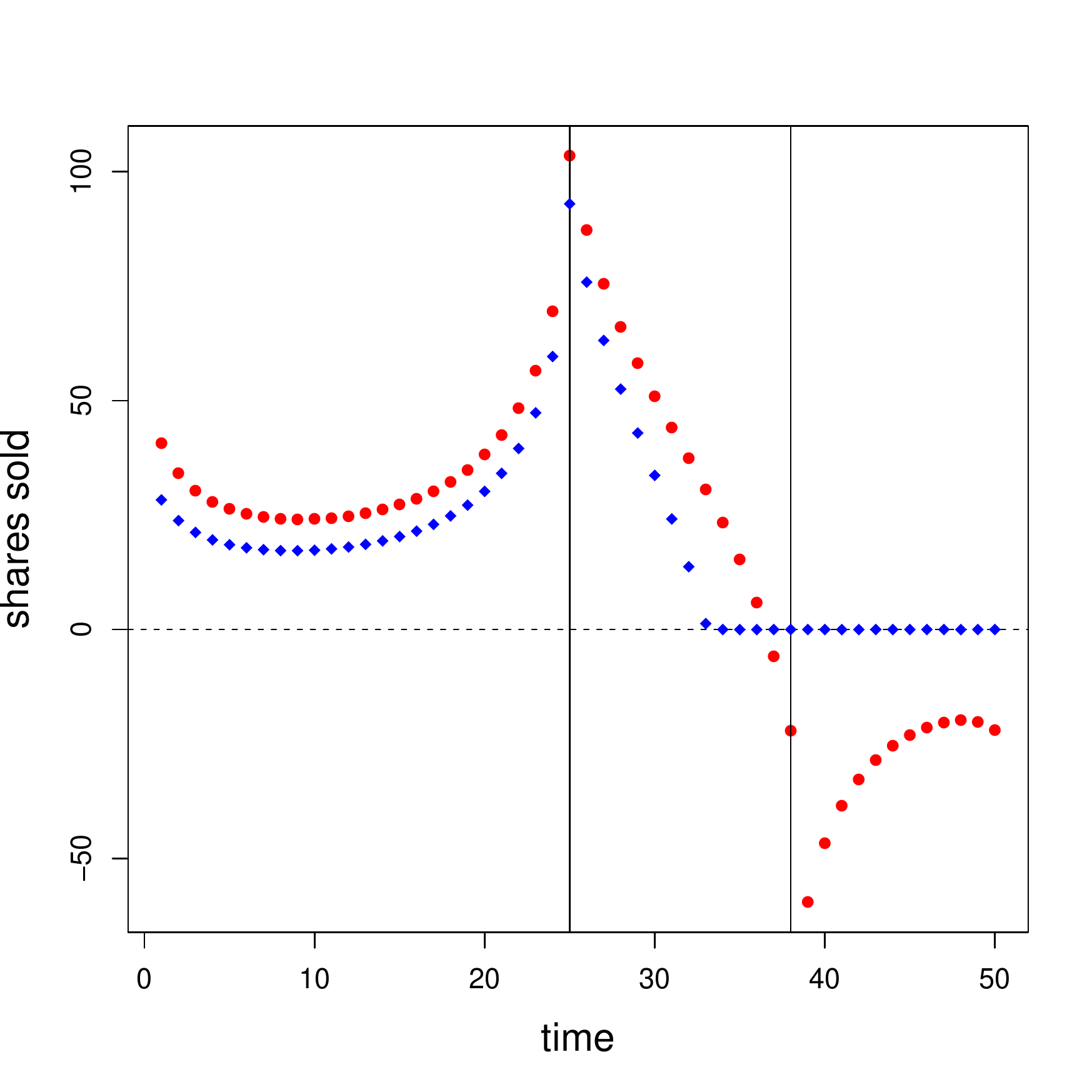}
	\caption{Optimal schedule without (red) and with (blue) constraint on trade sign for a VWAP with benchmark interval $T_1=25$ and $T_2=38$ (vertical lines).}
	\label{vwapt}
\end{figure}

We now come to the case of the benchmark period $[T_1,T_2]$ not coincident with the trading period $[0,T]$. Figure \ref{vwapt} shows the solution for drifless prices and risk neutral broker when $T_1=25=T/2$ and $T_2=38\simeq3T/4$. The figure shows the result with (black) and without (red) constraint on the sign of the trades. We observe that the optimal solution is to trade before, during, and after the benchmark interval. If the constraint that all the trades must have the same sign is imposed, it is optimal not to trade after the benchmark period. Interestingly, before the start of the benchmark period the trading pattern resembles the U-shape of the optimal execution under IS (see Eq. \ref{eq:gsssol} for the expression in continuous time), while during the benchmark period the trading pattern is similar to the one obtained when the trading interval coincides with the benchmark interval (see Fig. \ref{fig:vwapbase}).

\begin{figure} [t]
	\centering
	\includegraphics[width=0.45\textwidth]{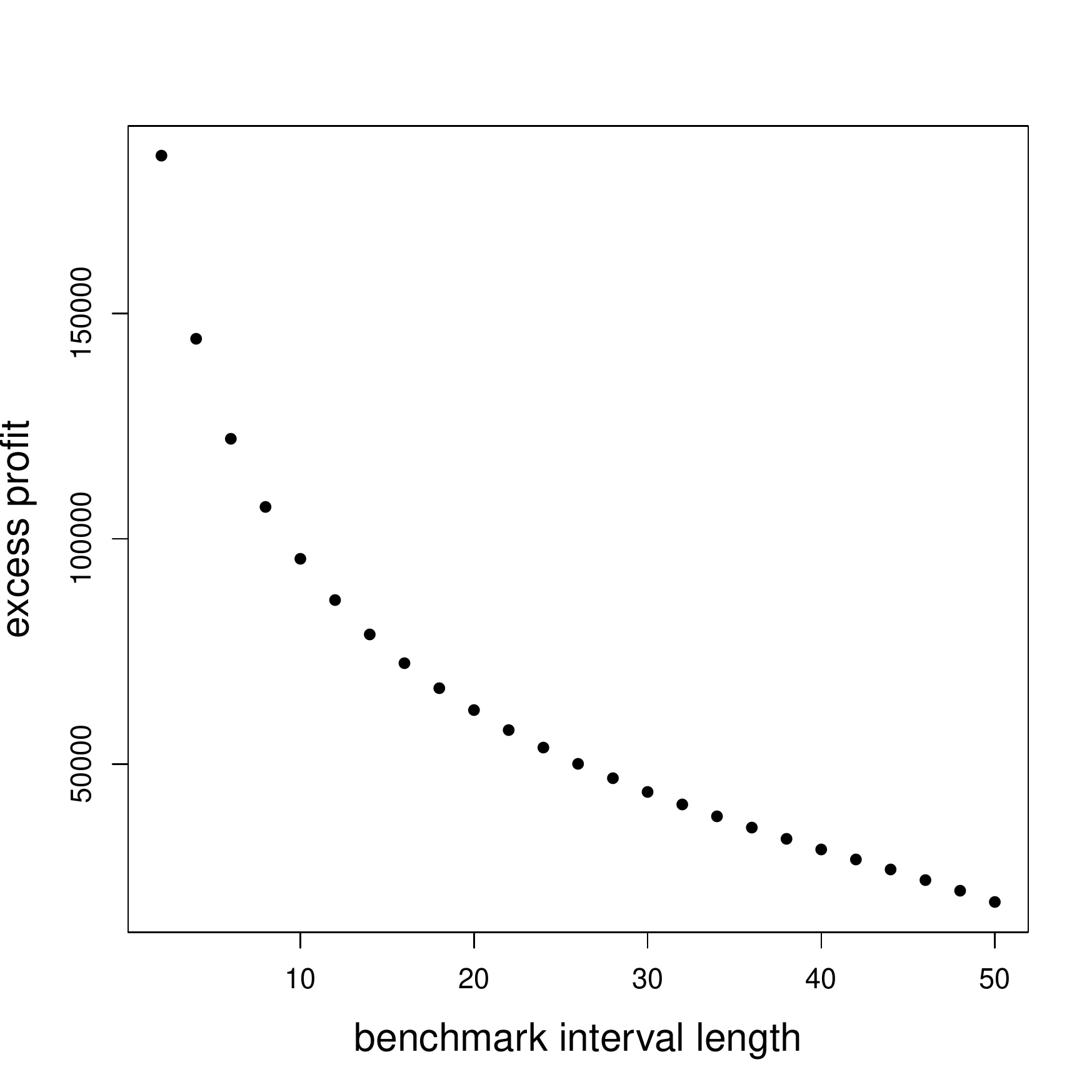}
	\includegraphics[width=0.45\textwidth]{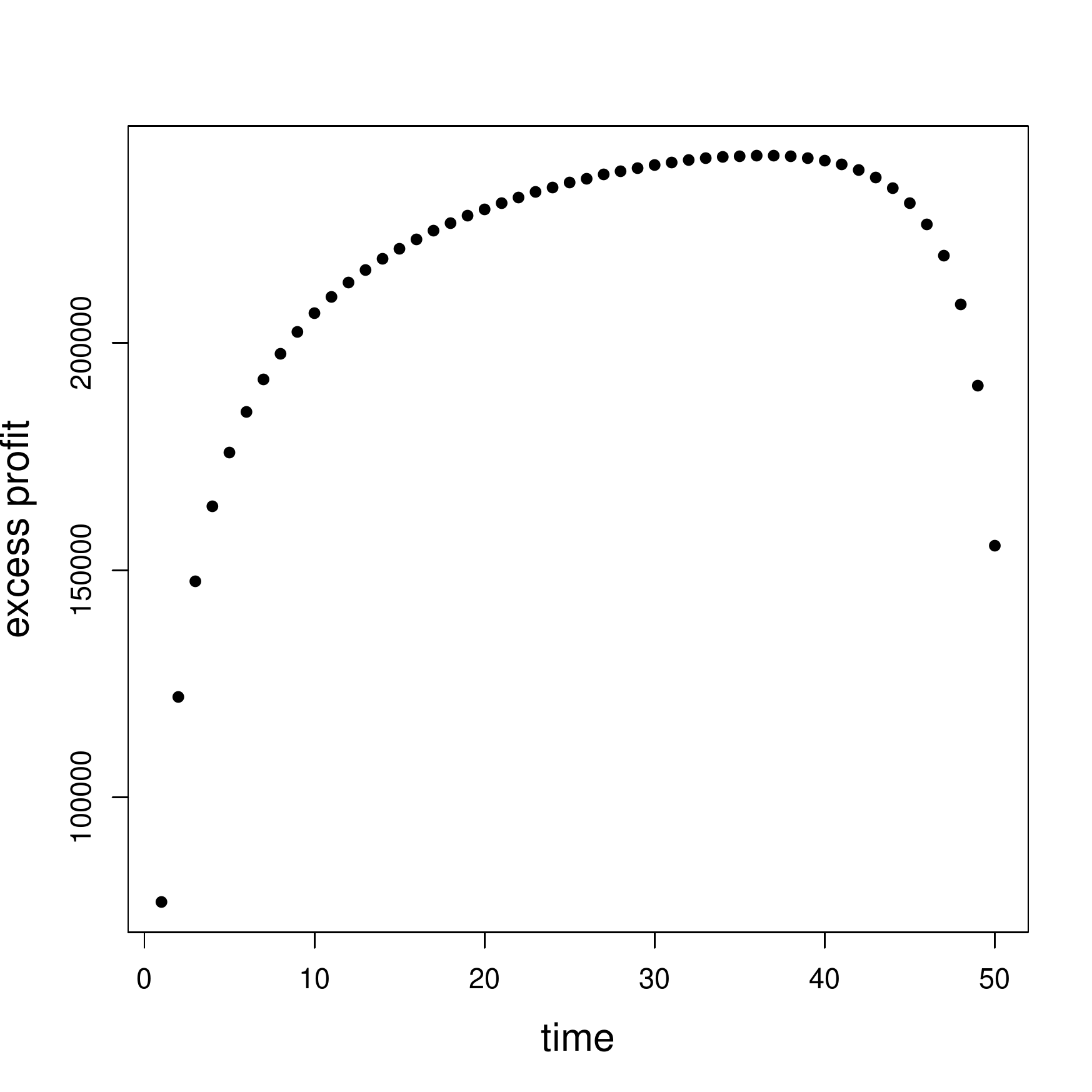}
	\caption{Excess profit of the broker for a VWAP execution with benchmark interval different from trading interval. The left panel shows the profit as a function of the length of the benchmark period when it is centered in $T/2$. The right panel shows the profit as  a function of the time within the trading period when the benchmark period has unit length.}
	\label{profit}
\end{figure}

Finally, we consider how the expected excess profit of the broker ${\mathbb E}[ ({\mathbf x}-x_0{\bm \eta})' {\mathbf S}]$ depends on the benchmark interval.  The excess profit is the difference between the cash at the end of the trading period and the VWAP in the benchmark period (which is the cash given by the broker to the client). We again consider $T=50$, drifless prices, risk neutral broker, and the other parameters as above. The left panel of figure \ref{profit} shows the expected excess profit of a benchmark period centered in $T/2$ and of variable length. It is clear that the interval providing the largest profit is the shortest one. Given this result, the right panel shows the excess profit for a benchmark period of length one as a function of the time within the trading period where the benchmark period is located. The shape is non-monotonic and, for the chosen parameter, the benchmark period providing the largest profit is of length one and located at time $T_1=T_2=36$. 

In conclusion the benchmark period providing the largest profit is very short and located in the second half of the trading interval. It is important to remark however that we have implicitly assumed that the market impact model of Eq. (\ref{eq:tim}) continues to hold also for the very large trading intensities required for short benchmark periods. This is unlikely in reality and additional constraints (for example on the maximal trading speed) should be added to the optimization of Proposition \ref{prop:discrete} to have more realistic results.

\subsection{Including executed volume in the benchmark}
Especially for short benchmarks, the volume coming from the optimal execution can be a significant fraction of the market volume and therefore one should add it to build the benchmark. Thus Eq. (\ref{bench}) can be imprecise and should be replaced by
$$
\eta_\ell=\frac{V_\ell+|x_\ell|}{\sum_{\ell\in B} V_\ell +\sum_{\ell\in B} |x_\ell|}I_{\ell\in B}
$$
Due to the fact that $x_\ell$ appears with absolute value and in the denominator, plugging this benchmark price in the optimization leads to a non quadratic optimization which can be very hard to solve. We will consider here the case when $|x_\ell | \ll V_\ell$, leading to the expansion
$$
\eta_\ell \approx\frac{V_\ell+x_\ell}{\sum_{\ell\in B} V_\ell}\left(1-\frac{\sum_{\ell\in B} x_\ell}{\sum_{\ell\in B} V_\ell }\right)I_{\ell\in B}
$$
Note that the $L_1$ norm of this vector, $\sum_\ell \eta_\ell$, is equal to one.
 For simplicity we consider the case $V_\ell=V$ and denote with $\Delta=T_2-T_1+1$ the length of the benchmark time, thus  
 $$
\eta_\ell \approx \frac{1}{\Delta}+\frac{x_\ell}{\Delta V}-\frac{\sum_{\ell\in B} x_\ell}{\Delta^2V}
 $$
 The argument of the utility function can be rewritten as ${\mathbf x}-x_0[{\mathbf a}+M {\mathbf x}]$ where $a_\ell= I_{\ell\in B}/\Delta $ and 
 $$
 M_{ij}=\frac{1}{V} I_{i\in B} I_{j\in B} \left(I_N-\frac{1}{\Delta} \hat 1\right)
$$
and $I_N$ is the $N \times N$ identity matrix and $\hat 1$ is the unit matrix (matrix where all elements are ones).
In conclusion the optimization is the same as before with the only change $G \to G- x_0 M$

%\iffalse 
\section{Conclusions} \label{sec:concl}

In this paper we have set and solved  the problem of the optimal execution of an order when the benchmark is the volume weighed average price on a specific time interval and the price impact is transient. 
By considering the general case when the trading interval is larger than the benchmark interval, we have shown that several existing optimal execution benchmarks (Implementation Shortfall, Target Close, VWAP, and TWAP) can be seen as special cases. 
We have considered the solution in continuous time, mapping the maximization problem into the solution of an integral equation, in line with what was done for IS in \cite{Gatheral2}.  
Solution in discrete time has been reduced to standard quadratic optimisation problem.
We have not explicitly considered transaction costs and child order placement as being part of optimisation which would be required for practical applications.
One of the ways to approximately take this into account is by incorporating tactical cost into the impact kernel \cite{Busseti}

%While in the risk neutral drifless case, transaction triggered price manipulation
%\fi
\newpage

\appendix

\section{Proofs of propositions}

\subsection{Proof of proposition 1}
Since $X_T-x_0VWAP_{T_1}^{T_2}$ is a Gaussian distributed variable, the broker maximizes
\begin{eqnarray}\label{eq:util}
U[{\mathbf x}]={\mathbb E}_0\left[\int_0^T (\dot x_t-x_0\eta_t)S_tdt\right]-\gamma{\mathbb V}_0\left[\int_0^T (\dot x_t-x_0\eta_t)S_tdt\right]
\end{eqnarray}
The expected value is
\begin{eqnarray}
&&{\mathbb E}_0\left[\int_0^T (\dot x_t-x_0\eta_t)S_tdt\right]=\\
&&{\mathbb E}_0\left[\int_0^T (\dot x_t-x_0\eta_t)S_0dt\right]+{\mathbb E}_0\left[\int_0^T (\dot x_t-x_0\eta_t)\int_0^t f(\dot x_s)G(t-s)dsdt\right]+ \nonumber \\
&&{\mathbb E}_0\left[\int_0^T (\dot x_t-x_0\eta_t)\int_0^t\sigma_s dW_s dt\right]=\nonumber \\
&&{\mathbb E}_0\left[\int_0^T (\dot x_t-x_0\eta_t)\int_0^t f(\dot x_s)G(t-s)dsdt\right] \nonumber
\end{eqnarray}
because the first term identically vanishes due to the normalization of $\dot x_t$ and $\eta_t$ and the third term is the expectation of a stochastic integral.

In the case of linear impact, this becomes
\begin{eqnarray}
-k\int_0^T\left(\dot x_t-x_0 \eta_t\right)\int_0^t \dot x_s G(t-s)dsdt=\nonumber \\
-k\left[\frac{1}{2}\int_0^T \int_0^T \dot x_t \dot x_s G(|t-s|)ds~dt-x_0\int_0^T \eta_t dt \int_0^t G(t-s) \dot x_sds\right]
\end{eqnarray}

Similarly for the variance term
\begin{eqnarray}
&&{\mathbb V}_0\left[\int_0^T (\dot x_t-x_0\eta_t)S_tdt\right]=\\
&&{\mathbb V}_0\left[\int_0^T (\dot x_t-x_0\eta_t)S_0dt\right]+{\mathbb V}_0\left[\int_0^T (\dot x_t-x_0\eta_t)\int_0^t f(\dot x_s)G(t-s)dsdt\right]+ \nonumber \\
&&{\mathbb V}_0\left[\int_0^T (\dot x_t-x_0\eta_t)\int_0^t\sigma_s dW_s dt\right]=\nonumber \\
&&{\mathbb V}_0\left[\int_0^T (\dot x_t-x_0\eta_t)\int_0^t\sigma_s dW_s dt\right]={\mathbb E}_0\left[\left(\int_0^T (\dot x_t-x_0\eta_t)\int_0^t\sigma_s dW_s dt\right)^2\right]\nonumber
\end{eqnarray}

The last expectation can be written as
\begin{eqnarray}
{\mathbb E}_0\left[\int_0^T\int_0^T dt dt' (\dot x_t-x_0\eta_t)(\dot x_{t'}-x_0\eta_{t'})\int_0^t\int_0^{t'} \sigma_s\sigma_{s'} dW_sdW_{s'}\right]=\nonumber \\
\int_0^T\int_0^T dt dt' (\dot x_t-x_0\eta_t)(\dot x_{t'}-x_0\eta_{t'})\int_0^t\int_0^{t'} \sigma_s\sigma_{s'}{\mathbb E}_0\left[dW_sdW_{s'}\right]=\nonumber \\
\int_0^T\int_0^T dt dt' (\dot x_t-x_0\eta_t)(\dot x_{t'}-x_0\eta_{t'})\int_0^t\int_0^{t'} \sigma_s\sigma_{s'}\delta(s-s')=\nonumber\\
\int_0^T\int_0^T dt dt' (\dot x_t-x_0\eta_t)(\dot x_{t'}-x_0\eta_{t'})\int_0^{t \wedge t'} \sigma_s^2 ds
\end{eqnarray}
Finally, given that $k>0$ and $\gamma >0$, the maximization of the utility $U[{\mathbf x}]$ is equivalent to the minimization of the functional $C[{\mathbf x}]$ of Eq. (\ref{eq:cost}).
\

\subsection{Proof of proposition 2}

The quantity to minimize in  a VWAP execution is

\begin{equation}\label{mineq}
C[{\mathbf x}]=\frac{1}{2}\int_0^T \int_0^T dx_t dx_s G(|t-s|)-\frac{x_0}{T}\int_0^T dt \int_0^t G(t-s) dx_s
\end{equation}
that we rewrite as $C[{\mathbf x}]=Q[{\mathbf x}]+K[{\mathbf x}]$. 

Following \cite{Gatheral2,LeHalle}, we consider a strategy 
\begin{equation}
dy_s=\delta_{t_2}(ds)-\delta_{t_1}(ds)~~~~~~~~~~0\le t_1 \le t_2 \le T
\end{equation}
and indicate with ${\mathbf x^*}$ the optimal strategy. Hence, setting ${\mathbf z}={\mathbf x^*}+\alpha {\mathbf y}$, it is
\begin{equation}
C[{\mathbf z}]=Q[{\mathbf x^*}]+\alpha^2Q[{\mathbf y}]+2\alpha Q[{\mathbf x^*},{\mathbf y}]+K[{\mathbf x^*}]+\alpha K[{\mathbf y}]
\end{equation}
where $Q[{\mathbf x},{\mathbf y}]=2^{-1}\int \int G(|t-s|)dx_s dy_t=Q[{\mathbf y},{\mathbf x}]$. The quantity
\begin{equation}
K[{\mathbf y}]=-\frac{x_0}{T}\left[\int_0^Tdt G(t-t_2)\theta(t-t_2)-\int_0^Tdt G(t-t_1)\theta(t-t_1)\right]
\end{equation}
where $\theta(x)$ is the step function, while %, as in \cite{Gatheral2,LeHalle}
\begin{equation}
Q[{\mathbf x},{\mathbf y}]=\frac{1}{2}\int_0^T G(|t_2-t|)dx_t-\frac{1}{2}\int_0^T G(|t_1-t|)dx_t
\end{equation}
If the strategy ${\mathbf x^*}$ is optimal then
\begin{equation}
\frac{\partial {\mathbb E}_0[C[{\mathbf z}]]}{\partial \alpha} \bigg |_{\alpha=0} =2{\mathbb E}_0[Q[{\mathbf x^*},{\mathbf y}]]+{\mathbb E}_0[K[{\mathbf y}]]=0
\end{equation}
i.e. if
\begin{equation}
\int_0^T G(|t-s|)dx_s^*-\frac{x_0}{T}\int_0^TdsG(s-t)\theta(s-t)=\lambda
\end{equation}
or equivalently
\begin{equation}
\int_0^T G(|t-s|)dx_s^*-\frac{x_0}{T}\int_t^TdsG(s-t)=\lambda
\end{equation}
 where $\lambda$ is a constant set by the normalization on the total volume traded
 \begin{equation}
 \int_0^T dx^*_s=x_0
 \end{equation}
 Generically we will be interested in the trading velocity defined in $dx^*_s=v_s ds$, thus we solve
 \begin{equation}
\int_0^T G(|t-s|)v_sds-\frac{x_0}{T}\int_t^TG(s-t)ds=\lambda~~~~~~~s.t. ~~\int_0^T v_s ds=x_0
\end{equation}

\subsection{Proof of proposition \ref{prop:discrete}}

The utility function is  ${\cal U}[({\mathbf x}-x_0{\bm \eta})'  {\mathbf S}]$.
Since everything is Gaussian and assuming as in the continuous time case a CARA utility function with risk aversion $2\gamma$, the expected utility is
$$
U[{\mathbf x}]={\mathbb E}_0[ ({\mathbf x}-x_0{\bm \eta})' {\mathbf S}] -\gamma {\mathbb V}_0[ ({\mathbf x}-x_0{\bm \eta})' {\mathbf S}]
$$
The mean value term is 
\begin{eqnarray*}
{\mathbb E}_0[ ({\mathbf x}-x_0{\bm  \eta})'  {\mathbf S}]=&&({\mathbf x}-x_0{\bm \eta})' {\mathbf 1} S_0 -k({\mathbf x}-x_0{\bm \eta})' G {\mathbf  x}+\tau^{1/2}({\mathbf  x}-x_0{\bm  \eta})' L {\bm \mu}=\\
&&-k({\mathbf x}-x_0{\bm \eta})' G {\mathbf x}+\tau^{1/2}({\mathbf x}-x_0{\bm \eta})' L {\bm \mu}
\end{eqnarray*}
where the first term vanishes because ${\mathbf  x}' {\mathbf 1}=x_0$ and ${\bm  \eta}' {\mathbf 1}=1$.

The variance term is
$$
{\mathbb V}_0[{\mathbf x}-x_0{\bm  \eta})'  {\mathbf S}]={\mathbb V}_0[k({\mathbf x}-x_0{\bm  \eta})' G {\mathbf  x} +\tau^{1/2}({\mathbf x}-x_0{\bm \eta})' L {\bm \epsilon}]=\tau {\mathbb V}_0[({\mathbf  x}-x_0{\bm  \eta})' L {\bm  \epsilon}]=
$$ 
$$
=\tau ({\mathbf x}-x_0{\bm  \eta})' L\Sigma L'({\mathbf x}-x_0{\bm \eta})
$$
Thus the maximization of the expected utility is equivalent to the minimization of
$$
k({\mathbf  x}-x_0{\bm \eta})' G {\mathbf x}-\tau^{1/2}({\mathbf x}-x_0{\bm \eta})' L {\bm \mu} +\gamma\tau ({\mathbf x}-x_0{\bm \eta})' L\Sigma L'({\mathbf x}-x_0{\bm \eta})
$$
which can me rewritten in matrix form as
$$
{\mathbf x}' A {\mathbf x} - {\mathbf b}' {\mathbf x}+C
$$
where $A$ and ${\mathbf b}$ are given in Eq. (\ref{eq:param1}) and (\ref{eq:param2}) and
$$
C=\tau^{1/2}x_0{\mathbf \eta}' L {\mathbf \mu}+\gamma\tau x_0^2 {\mathbf \eta}' L\Sigma L'{\mathbf \eta}
$$
is a constant not affecting the optimal solution (but of course affecting the value of the optimal expected utility).

To prove that $A$ is Positive Definite (PD), let us note first that $G$ is PD, since it is lower triangular with diagonal elements $G_{ii}=G(0)>0$. The other term, $L\Sigma L'$ can be rewritten as $BB'$, where $B=LS$ and $S$ is the lower triangular matrix obtained from the Cholesky decomposition of $\Sigma$ (which exists because $\Sigma$ is PD). Clearly $B$ is lower triangular with the same diagonal entries of $S$. These entries are positive, again because $\Sigma$ is PD, and therefore  $B$ is invertible. The product of an invertible matrix and its transpose, such as  $L\Sigma L'$, is PD. Finally, $A$ is the sum of two PD matrices, and therefore it is PD.

  \end{document}